\documentclass[conference]{IEEEtran}

\usepackage{amsmath}
\usepackage{amssymb}
\usepackage{algorithmic}
\usepackage{algorithm}
\usepackage{booktabs}
\usepackage{graphicx}
\usepackage{multirow}
\usepackage[caption=false,font=footnotesize]{subfig}
\usepackage{enumitem}
\usepackage{xcolor}
\usepackage{url}

\providecommand{\Description}[1]{}
\makeatletter
\renewcommand{\@IEEEfigurecaptionsepspace}{\vskip0pt}
\makeatother

\begin{document}

\title{HBFWeave: Coordinating DRAM and High Bandwidth Flash
for Sparse LLM Inference}

\author{
\IEEEauthorblockN{\small Tenghao Zhao\textsuperscript{1}, Shuzhang Zhong\textsuperscript{1}, HongKai Liao\textsuperscript{1}, Zizhuo Fu\textsuperscript{1}, Tongbin Zhao\textsuperscript{1}, Yifan Zhou\textsuperscript{1}, Weikai Xu\textsuperscript{2}, Meng Li\textsuperscript{1}}
\IEEEauthorblockA{\textsuperscript{1}School of Integrated Circuits, Peking University, Beijing, China\\
\textsuperscript{2}The Hong Kong University of Science and Technology, Hong Kong}
}

\maketitle

\begin{abstract}
Long-context sparse attention reduces computation and read traffic, but the full KV cache must still be stored. High-bandwidth Flash (HBF) provides the needed capacity, but irregular sparse KV accesses can waste data in Flash page reads and cause plane imbalance. Continuous KV writes can stall reads during Flash program and stress endurance. To address these challenges, we present HBFWeave, a mixed DRAM--HBF system. It replaces a few Flash dies with DRAM to cache frequently reused KV pages while retaining most Flash capacity and read bandwidth. HBFWeave uses reuse-aware admission to reduce Flash reads within this cache budget and ready-plane scheduling to serve residual misses without waiting behind busy planes. Read-prioritized program waves and request-grouped blocks further limit write interference and garbage-collection (GC) copying. Across four sparse models and 16 workloads, HBFWeave achieves an average throughput speedup of 2.65$\times$ and P50/P90 TPOT speedups of 2.55$\times$/3.81$\times$ over HBF-only.
\end{abstract}


\section{Introduction}
\label{sec:introduction}

As the context lengths of large language models (LLMs) grow, the KV cache imposes increasing demands on memory capacity and bandwidth. Recent models have been trained with sparse attention to reduce KV-cache read bandwidth demands during long-context inference~\cite{GLM53FlashFrontier, lai2026MiniMaxSparse, deepseek-ai2026DeepSeekV4Highly, qiu2026DesignQwen38Next}. This creates a capacity--bandwidth mismatch: the KV cache grows with context length, while each decoding step reads only a small fraction of it. Keeping the entire cache in high-bandwidth memory (HBM) therefore consumes scarce capacity without a proportional need for bandwidth~\cite{xie2026HiSparseScaling, jonnalagadda2026SPLASHCoDesigning}. This motivates placing sparse KV in a larger-capacity memory tier that serves irregular reads efficiently and provides sufficient bandwidth.

\begin{figure}[!t]
  \centering
  \includegraphics[width=\columnwidth]{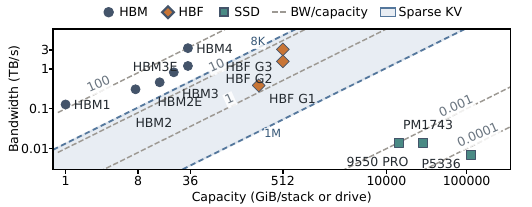}
  \caption{HBM/HBF stacks and solid-state drives (SSDs): capacity versus read bandwidth (SSD: peak
  sequential). Blue: illustrative sparse-KV demand for 8K--1M at 20-ms TPOT
  and 2,048 selections/layer; gray: constant bandwidth/capacity.}
  \label{fig:memory-capacity-bandwidth}
  \Description{Log-scale scatterplot of capacity versus read bandwidth.
  HBM stacks have low capacity and high bandwidth, HBF stacks offer hundreds
  of GiB with high bandwidth, and SSDs offer much higher capacity with lower
  sequential read bandwidth. Gray diagonals mark constant bandwidth per unit
  capacity; a blue band spans sparse-KV read demand for 8K--1M context.}
\end{figure}

High-bandwidth flash (HBF) offers a large-capacity tier with high aggregate read bandwidth, but serves data through page-granular NAND operations and parallel Flash planes~\cite{jonnalagadda2026SPLASHCoDesigning}. We define the bandwidth-to-capacity ratio as peak read bandwidth divided by storage capacity (GB/s/GiB). The dashed lines in Fig.~\ref{fig:memory-capacity-bandwidth} mark constant values of this ratio. Compared with HBM, HBF provides greater capacity (256--512 GiB per stack) but a lower bandwidth-to-capacity ratio (1.5--6 GB/s/GiB)~\cite{agrawalHBFHotChips2026}. The blue band illustrates how sparse-KV logical demand falls with context length and overlaps HBF's range, motivating HBF as a capacity tier for sparse attention. However, irregular sparse KV reads can fetch unused data from Flash pages and create uneven plane loads, reducing effective HBF bandwidth. Meanwhile, continuous KV-cache writes and garbage collection (GC) accelerate wear and shorten Flash lifetime. They also compete with foreground KV reads during concurrent serving. Thus, HBF-backed sparse attention must manage irregular reads and writes to sustain serving performance and lifetime.

Recent systems have explored HBF-backed sparse attention. SparseTier keeps the index keys in HBM and the KV cache in HBF~\cite{lee2026EfficientLongContext}, while SPLASH retains a recent token KV cache window in HBM for each request and moves older KV cache to HBF~\cite{jonnalagadda2026SPLASHCoDesigning}. These studies establish HBF as a viable capacity tier for long-context inference. Both evaluate daisy-chained HBM and HBF stacks. Only a small subset of the HBF-resident KV cache is repeatedly accessed, so dedicating a full HBM stack to caching these pages may be excessive. A promising option is to replace a few Flash dies within each HBF stack with DRAM to cache frequently reused KV pages and avoid repeated Flash reads. But this replacement will reduce Flash capacity and bandwidth, so the number of DRAM dies must balance read savings against these losses.

To realize this approach, we present HBFWeave, a mixed DRAM--HBF system that replaces only a few Flash dies with DRAM. Flash stores the full sparse KV cache and index keys, while DRAM caches hot KV pages. To use this limited DRAM effectively and control residual Flash reads and write/GC interference, HBFWeave combines reuse-aware admission, ready-plane scheduling, and read-prioritized program waves with request-grouped blocks. Our contributions are:

\begin{enumerate}[label=(\arabic*),leftmargin=1.6em,labelsep=0.4em,itemsep=1pt,topsep=2pt]
  \item We design a mixed DRAM--HBF system for disaggregated sparse-attention serving. Flash stores the full sparse KV cache and index keys; DRAM caches reused KV pages.
  \item We propose three mechanisms. Reuse-aware admission reduces Flash reads. Ready-plane scheduling exploits available planes. Read-prioritized program waves and request-grouped blocks limit read stalls and GC copying.
  \item Across four models and 16 cases, we achieve a 2.65$\times$ average throughput speedup, 2.55$\times$/3.81$\times$ P50/P90 TPOT speedups, and 1.71--3.88$\times$ lifetime gains over HBF-only.
\end{enumerate}

\section{Background and Motivation}\label{sec:background}
\subsection{Sparse LLM Workload Characteristics}\label{sec:sparse-workloads}
\begin{figure}[!t]
  \centering
  \includegraphics[width=\columnwidth]{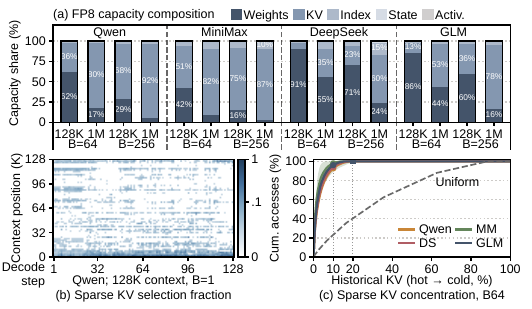}
  \caption{Sparse LLM workloads: (a) FP8 capacity; (b) Qwen KV selections; (c) KV access skew (DS: DeepSeek; MM: MiniMax).}
  \label{fig:sparse-workload}
  \Description{The top panel shows FP8 resident capacity shares across four sparse LLMs, two batch sizes, and two context lengths. The lower panels show sparse KV selection over 128 decoding steps in one Qwen request and the concentration of KV accesses across four models at batch 64.}
\end{figure}

We study four representative large-scale mixture-of-experts (MoE) LLMs with sparse or hybrid attention, as summarized in Table~\ref{tab:sparse-models}. Fig.~\ref{fig:sparse-workload}(a) shows how resident capacity shifts with context length and batch size. From batch size $B=64$ at 128K to $B=256$ at 1M, weight share falls from 42--91\% to 2--24\%, while KV cache share rises from 7--51\% to 60--92\%. Across all cases, full-resolution index keys occupy approximately 0.8--15\% of resident capacity, with their size growing with batch size and context length. FP8 linear-attention states occupy 1.69--8.50 GiB in Qwen and GLM at \(B=64\)--256; their size grows with batch size but not with context length.

Fig.~\ref{fig:sparse-workload}(b) illustrates sparse KV selection in a representative Qwen trace: each step reads a small, scattered fraction of the cache, and some context regions are revisited. Fig.~\ref{fig:sparse-workload}(c) shows strongly skewed access frequencies across the four models at \(B=64\) and 128K (median and interquartile range): the hottest 10\% of KV entries account for 92.7--98.0\% of accesses. HBF stores the full sparse KV cache, but scattered selections waste page-read bandwidth and repeatedly incur long Flash sensing latency. This reuse creates an opportunity to cache hot KV pages in DRAM and reduce rereads from Flash.

\begin{table}[!t]
  \centering
  \caption{Sparse LLM configurations and FP8 KV units.}
  \label{tab:sparse-models}
  \fontsize{7.5}{9}\selectfont
  \setlength{\tabcolsep}{1.5pt}
  \begin{tabular}{@{}p{0.32\columnwidth}
                      p{0.13\columnwidth}
                      p{0.38\columnwidth}
                      p{0.12\columnwidth}@{}}
    \toprule
    Model & Params(B) & Attention layers & KV (FP8) \\
    \midrule
    \raggedright DeepSeek-V4-Flash~\cite{deepseek-ai2026DeepSeekV4Highly}
      & $284/13$
      & \raggedright 21 sparse, 20 dense, 2 local
      & \raggedright 512~B \tabularnewline
    \addlinespace[1pt]

    \raggedright GLM-5.3-Flash~\cite{GLM53FlashFrontier}
      & $320/18$
      & \raggedright 34 linear, 11 sparse
      & \raggedright 2~KiB \tabularnewline
    \addlinespace[1pt]

    \raggedright Qwen3.8-Flash-Next~\cite{qiu2026DesignQwen38Next}
      & $125^{\dagger}/6$
      & \raggedright 36 linear, 12 sparse
      & \raggedright 4~KiB \tabularnewline
    \addlinespace[1pt]

    \raggedright MiniMax-M3~\cite{lai2026MiniMaxSparse}
      & $428/23$
      & \raggedright 3 dense, 57 sparse
      & \raggedright 32~KiB \tabularnewline
    \bottomrule
  \end{tabular}
  \par\smallskip
  \begin{minipage}{\columnwidth}
    \scriptsize\raggedright
    Params: total/active; $\dagger$ excludes 51B embeddings
    and 4B multi-token predictor. KV selection: compressed entry (DeepSeek), 4-token group
    (Qwen/GLM), 128-token block/head (MiniMax). Index key: 128~B per 4 tokens except MiniMax (per token).
  \end{minipage}
\end{table}

\subsection{HBF Characteristics and Mixed-Stack Tradeoffs}
\label{sec:hbf-integration}
High Bandwidth Flash (HBF) stacks NAND Flash dies above a logic base die using through-silicon vias (TSVs)~\cite{ballapuramHBFBaseDie2026,agrawalHBFHotChips2026,ha2026H3Hybrid}. This organization offers high storage density. Unlike HBM's fine-grained, nanosecond-scale accesses, NAND reads sense kilobyte-scale pages and take microseconds. HBF achieves high aggregate read bandwidth by accessing many NAND planes in parallel~\cite{zhong2026HBFlexFlexible,jonnalagadda2026SPLASHCoDesigning}. Programming Flash pages incurs substantial latency, while reclaiming space requires block erases and may involve copying live pages. Limited program/erase (P/E) endurance makes frequent updates costly~\cite{oliveira2026FLINTEfficientlyb,li2026HBFSucks,zhong2026HBFlexFlexible}. Together, these Flash costs constrain HBF's effectiveness for sparse-attention serving.

\begin{figure}[!t]
  \centering
  \includegraphics[width=\columnwidth]{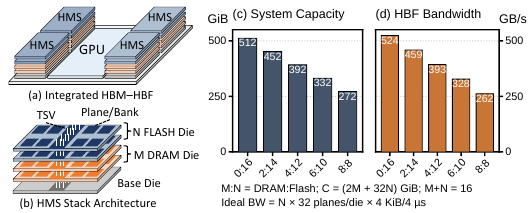}
  \caption{Mixed DRAM--HBF stack and allocation tradeoff. (a)--(b) Layout; (c) Capacity with 2 GiB/DRAM die and 32 GiB/Flash die; (d) HBF bandwidth.}
  \label{fig:hms-tradeoff}
  \Description{An integrated GPU memory system and a mixed DRAM--Flash stack, followed by bar charts showing decreasing raw capacity and ideal Flash array bandwidth as DRAM dies replace Flash dies in a 16-memory-die stack.}
\end{figure}

To reduce repeated Flash page sensing, we consider a heterogeneous memory stack (HMS) that places DRAM and Flash dies in the same stack~\cite{yin2026PotentialApplications,zhong2026HBFlexFlexible}. For sparse attention, Flash can store the complete KV cache and index keys, while DRAM caches hot KV pages selected repeatedly across decoding steps. Cache hits avoid repeated Flash page sensing, but DRAM and Flash compete for space within a fixed die budget. Fig.~\ref{fig:hms-tradeoff} illustrates the resulting capacity and ideal Flash array-bandwidth tradeoff for a 16-die stack. The modeled configuration assumes 32 planes per Flash die, 4 KiB pages, and 4~$\mu$s sensing. Replacing Flash dies with DRAM reduces both Flash capacity and bandwidth. These competing effects favor allocating only enough DRAM to capture KV reuse while retaining most Flash capacity and read bandwidth.

\subsection{Design Challenges for Mixed DRAM--HBF Stacks}\label{sec:Implications-for-HMS-Design}
\begin{figure}[!t]
  \centering
  \includegraphics[width=\columnwidth]{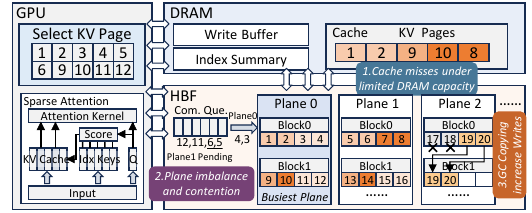}
  \caption{Three challenges in a mixed DRAM--HBF stack: limited cache capacity, plane imbalance and contention, and GC-induced write amplification.}
  \label{fig:motivation-mechanism}
  \Description{A GPU reads sparse KV pages through a DRAM cache and HBF. The diagram illustrates cache misses, queued reads to imbalanced Flash planes, and live-page copying during garbage collection.}
\end{figure}

Preserving Flash capacity and read bandwidth limits the number of DRAM dies in each stack. Fig.~\ref{fig:motivation-mechanism} illustrates three challenges facing this system: Limited DRAM capacity causes frequent Flash reads. Residual misses skew plane loads and contend for controller resources. Continuous KV and index-key writes delay reads; blocks mixing request lifetimes increase GC copying. We evaluate the four models in Table~\ref{tab:sparse-models} at batch size 256 and 128K context to quantify these effects.

\begin{figure}[!t]
  \centering
  \includegraphics[width=\columnwidth]{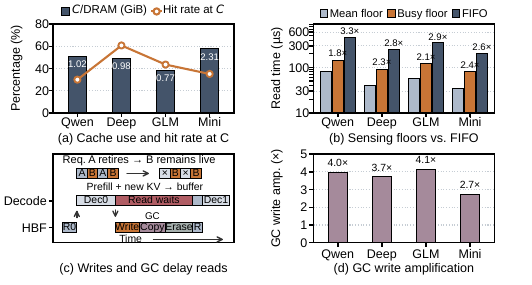}
  \caption{Mixed DRAM--HBF challenges: (a) cache occupancy and hit rate at $C$; (b) mean-plane and busiest-plane sensing floors versus FIFO read time (log scale), showing busiest/mean and FIFO/busiest ratios; (c) write/GC stalls; (d) random-placement GC write amplification on all-HBF stacks.}
  \label{fig:motivation-challenges}
  \Description{Four panels show cache hit rate and the cache target's share of DRAM, mean and busiest plane sensing-time floors against FIFO service time, an illustrative decode and HBF timeline in which writing, GC copying, and erasure delay a read, and GC write amplification across four sparse LLMs.}
\end{figure}

\noindent{\bf Challenge 1: Cache misses under limited DRAM capacity.}
Let $C$ denote the workload-wide capacity needed to hold all KV pages selected across requests and sparse-attention layers in one decode step. For the evaluated $B=256$, 128K cases, Qwen, DeepSeek, and GLM each use one GPU with 15 Flash and one 2~GiB DRAM die per stack; MiniMax uses two GPUs with 14 Flash and two 2~GiB DRAM dies per stack. Fig.~\ref{fig:motivation-challenges}(a) shows that each stack's share of $C$ occupies 0.77--2.31~GiB, or 38--58\% of its DRAM, yet hits only 29.9--60.8\% of KV page references. Consequently, Flash page senses with caching remain 39--70\% of those without. The remaining DRAM needs to hold index summaries and stage incoming KV and index-key writes. Enlarging the cache shrinks this reserve or requires replacing a Flash die with DRAM, reducing HBF capacity and parallel read/write bandwidth. The challenge is to raise the cache hit rate within this die budget and reduce Flash reads.

\noindent{\bf Challenge 2: Plane imbalance and read contention.}
After DRAM hits, residual sparse KV misses create uneven plane loads. In Fig.~\ref{fig:motivation-challenges}(b), the busiest plane's sensing time is 1.8--2.4$\times$ the mean-plane floor. FIFO read service takes another 2.6--3.3$\times$ longer than the busiest-plane floor: requests to busy planes delay reads to ready planes, while finite command and transfer resources add contention. The challenge is to sustain parallel reads despite skewed misses and finite resources.

\noindent{\bf Challenge 3: Write interference and GC amplification.}
Decode reads only selected sparse KV pages, but prefill and decode write every generated sparse KV and index-key page to Flash. Fig.~\ref{fig:motivation-challenges}(c) shows read stalls from Flash programming and GC. Mixed-lifetime blocks increase write amplification because GC must copy live pages before erasing them. In a separate 10,000-request all-HBF stress case, random block placement yields 2.7--4.1$\times$ write amplification (Fig.~\ref{fig:motivation-challenges}(d)). The challenge is to limit write-induced read stalls and GC copying.

\section{HBFWeave Design}\label{sec:design}
To address the challenges in Section~\ref{sec:Implications-for-HMS-Design}, HBFWeave combines a mixed DRAM--HBF stack with reuse-aware KV caching, ready-plane scheduling for residual Flash misses, and read-prioritized program waves with request-grouped block allocation. We next detail the architecture and methods.

\subsection{System Architecture and Data Placement}\label{sec:architecture}
\begin{figure*}[!t]
  \centering
  \includegraphics[width=\textwidth]{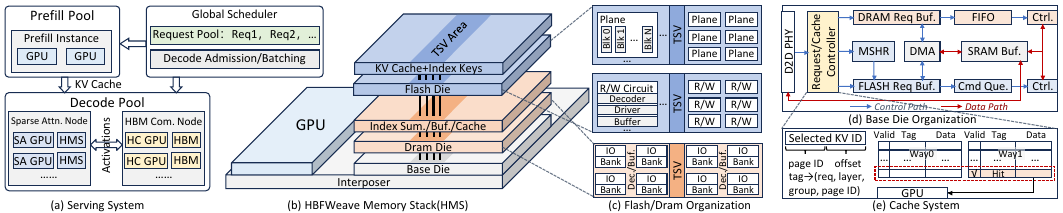}
  \caption{HBFWeave system architecture and KV-page cache. (a) Serving pools with sparse-attention and HBM compute nodes; (b) GPU and mixed DRAM--HBF stack; (c) Flash and DRAM organization; (d) base-die control and data paths; (e) KV-page mapping and cache lookup.}
  \label{fig:hbfweave-architecture}
  \Description{The serving system connects sparse-attention and HBM compute GPUs. Each mixed stack
  contains Flash and DRAM dies above a base die. The base die includes request
  routing, DRAM and Flash controllers, an MSHR, DMA, and an SRAM buffer. The
  DRAM KV-page cache stores data with corresponding valid bits and tags; a
  selected KV ID is resolved to a page ID and offset for cache lookup.}
\end{figure*}

Fig.~\ref{fig:sparse-workload}(a) shows that KV cache can dominate capacity as context grows. Resident data of size \(M\) with per-step reads \(R\) needs \(R/(M\tau)\) bandwidth per byte to achieve TPOT \(\tau\). The four models read about 1.5\% of their resident sparse KV cache per step at 128K. This requires about 0.8~GB/s/GiB of logical bandwidth at a 20ms TPOT. DeepSeek's heavily compressed attention (HCA) KV, the linear-attention states in Qwen and GLM, and MiniMax's full-attention KV cache are read in full and require about 54~GB/s/GiB; MoE weights also demand more bandwidth per byte. Given HBF's capacity and HBM's bandwidth-to-capacity advantage (Fig.~\ref{fig:memory-capacity-bandwidth}), we use attention--FFN disaggregation~\cite{zhu2025MegaScaleInferEfficient} to place sparse KV in mixed DRAM--HBF stacks on sparse-attention (SA) nodes, while HBM compute (HC) nodes hold the full-read states and MoE weights (Fig.~\ref{fig:hbfweave-architecture}(a)). Each SA node connects to six 16-die DRAM--HBF stacks via die-to-die (D2D) links (Fig.~\ref{fig:hbfweave-architecture}(b)--(c)). DRAM and Flash share TSVs to reach the base die. HBF stores the full sparse KV cache and index keys; DRAM holds hot KV pages, index summaries, and write buffers.

Fig.~\ref{fig:hbfweave-architecture}(d)--(e) shows how a selected KV ID identifies a model-specific compressed entry or token block. This unit maps to one or more 4~KiB logical pages and in-page offsets. A page key combines request, layer, KV group, and logical page ID. The base-die request/cache controller checks DRAM cache tags and routes hits to the DRAM controller. A miss status holding register (MSHR) entry tracks each miss until it completes, while its read passes through the Flash request buffer and channel command queue to the target plane. The selected KV bytes are staged in base-die SRAM for GPU delivery. For admitted pages, the DMA engine transfers the sensed Flash page into DRAM within the stack.

\begin{figure}[!t]
  \centering
  \includegraphics[width=\columnwidth]{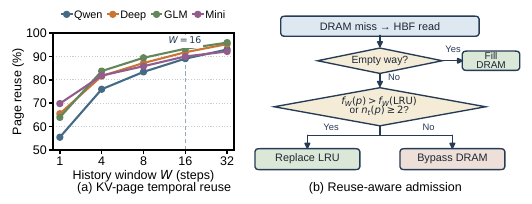}
  \caption{(a) Reuse at $B=256$/128K; dashed: $W=16$. (b) Admission.}
  \label{fig:cache-reuse-admission}
  \Description{Left: four curves show median KV-page reuse increasing with history window length; a dashed line marks the chosen 16-step window. Right: a flowchart shows that an HBF-read page fills an empty DRAM way, replaces the LRU page when its recent count is higher or it is requested twice in the current step, or bypasses DRAM.}
\end{figure}

\subsection{Reuse-Aware KV Cache Admission}
With limited DRAM capacity, raising cache hit rate requires favoring KV pages likely to be reused. Fig.~\ref{fig:cache-reuse-admission}(a) shows median KV-page reuse increasing as \(W\) grows from 1 to 32; we choose \(W=16\) because it reaches 89.1--93.3\% with little further gain (\(B=256\), 128K). This temporal reuse suggests that recent selection history can guide admission and improve hit rate within the limited cache budget. In addition, a 4~KiB cache line may hold eight 512~B DeepSeek entries or two 2~KiB GLM groups, so different selections within a layer can access the same Flash page. HBFWeave therefore tracks current-layer page references to enable second-reference admission.

Fig.~\ref{fig:cache-reuse-admission}(b) bases admission on two counts for page \(p\): \(f_W(p)\) counts the preceding \(W\) decode steps in which \(p\) was selected, and \(n_t(p)\) counts its current-layer references. A DRAM metadata table derives \(f_W(p)\) from a \(W\)-bit history and last-update step; base-die SRAM tracks \(n_t(p)\) and resets it after each layer. Each stack uses a set-associative cache indexed by the page key. Hits update least-recently-used (LRU) state. An HBF-read page fills an empty way. In a full set, it replaces the LRU victim only if its \(f_W\) is higher~\cite{einziger2017TinyLFUHighly} or \(n_t(p)\geq 2\).

\subsection{Ready-Plane Read Scheduling}
Although KV pages stripe round-robin across Flash planes, sparse selections and DRAM hits leave uneven residual misses. A FIFO issuer may wait for a busy plane while other planes can accept reads. HBFWeave's base-die controller tracks pending reads by target plane and each plane's availability. Subject to command-channel and outstanding-request limits, it selects the ready plane with the most pending reads. If none is ready, it waits for the earliest available plane to accept another read command, after which it chooses again among pending reads.

In Fig.~\ref{fig:read-write-mechanisms}, Plane~0 is sensing p7, with p8 waiting behind it. Planes~1 and 2 are ready; Plane~1 has two pending pages, p12 and p14, versus p9 on Plane~2. The controller issues the read for p12 while Plane~0 remains busy. Thus, reads to available planes proceed without waiting behind earlier requests to busy planes, making better use of plane parallelism without changing the selected KV IDs or their physical locations.

\subsection{Write Scheduling and Request Grouping}
\begin{figure}[!t]
  \centering
  \includegraphics[width=\columnwidth]{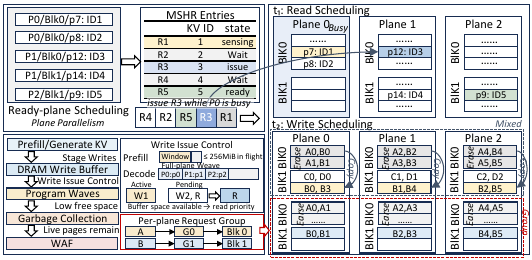}
  \caption{HBFWeave mechanisms. Top: ready-plane reads; bottom: prefill \mbox{window}, read-prioritized program waves, and request-grouped Flash blocks.}
  \label{fig:read-write-mechanisms}
  \Description{Top: KV IDs map to pages on busy Plane 0 and ready Planes 1 and 2; the scheduler issues p12 to Plane 1. Bottom: prefill and decode writes enter a DRAM buffer before parallel plane programming. Mixed A and B pages require copying live B pages when A retires, whereas request-grouped A pages permit direct block erasure.}
\end{figure}

Prefill transfers initial KV-cache and index-key pages in chunks, while decode generates new pages. Both streams enter the DRAM buffer. The two page types occupy separate Flash erase blocks and are striped round-robin across planes. In Fig.~\ref{fig:read-write-mechanisms} (bottom), staged pages form program waves of at most one page per plane; request completion or buffer pressure may trigger a partial wave. Each SA GPU caps outstanding prefill writes at 256~MiB; additional writes wait in DRAM. Larger windows permit more outstanding writes but may delay reads; smaller ones hold new writes in DRAM until earlier programs finish and may reduce throughput. When buffer space permits, reads bypass queued programs but wait for active ones.

Flash erases whole blocks within each plane. In Fig.~\ref{fig:read-write-mechanisms}, interleaving pages from requests A and B in one block leaves B pages live when A retires; GC must copy them before erasing that block. If A's pages occupy a separate block, it becomes fully invalid at retirement and can be erased without copying. HBFWeave maps each request to one of a bounded number of write groups per plane. A maps to G0 and B to G1, so their pages enter different open blocks. Each group opens a new block when its current one fills. Requests sharing a group may mix, but grouping reduces GC copying and write amplification without dedicating an open block per request.

\section{Evaluation}
\label{sec:evaluation}

\begin{figure*}[!t]
  \centering
  \includegraphics[width=\textwidth]{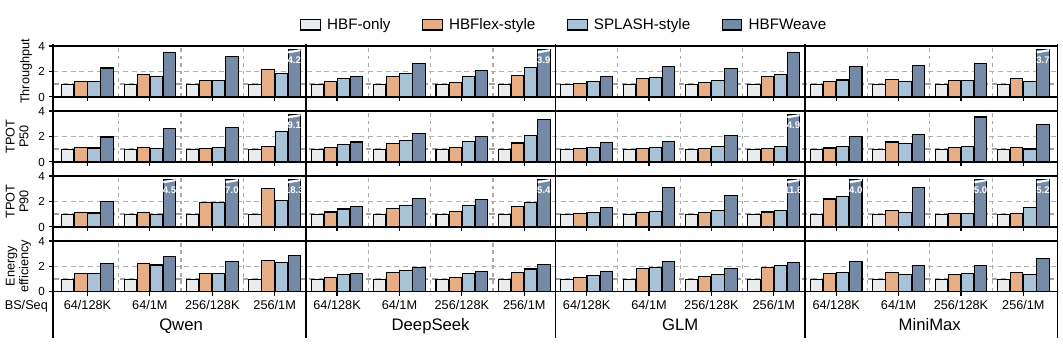}
  \caption{Throughput, TPOT P50/P90, and estimated energy-efficiency gains over HBF-only (BS/Seq: batch/context).}
  \label{fig:e2e-comparison}
\end{figure*}

\subsection{Experimental Setup}
\subsubsection{Workloads and System Configuration}
We evaluate the four models in Table~\ref{tab:sparse-models} with FP8, $B=64/256$, and 128K/1M-token contexts. Sparse-selection profiles come from LongBench v2~\cite{bai2025LongBenchV2a}. For 10,000 simulated requests, we repeat a LongBench-derived read profile and sample output lengths with a fixed seed. Requests become ready after prefill at 100 per second; arrivals pause when the upstream queue fills. We use a trace-driven simulator of mixed DRAM--HBF stacks, using HBFSim~\cite{li2026hbfsim} to model HBF and Ramulator~\cite{luo2024Ramulator20} to model DRAM. Table~\ref{tab:system-configuration} summarizes hardware parameters. One 2~GiB DRAM die retains 16 channels and 32 bank units per channel. Ramulator estimates 640~GB/s for its sequential 4~KiB reads, above the shared 512~GB/s TSV limit. Table~\ref{tab:case-resources} reports each stack's DRAM budget for the KV cache target \(C\), coarse index summaries~\cite{xu2026HISAEfficienta} (one key per index block), reserve (mainly cache tags and history metadata), and write buffer (WB). We estimate GPU compute energy using 0.52~pJ/FLOP~\cite{vartziotis2026TokensWatthours,antepara2025BenchmarkdrivenModelsa} and 2.6~pJ/bit for DRAM access~\cite{green2026PARE}. Flash read/program/erase are 8/100/10~pJ/bit~\cite{yanagidaira20251Tb3b,wang2026FlashAccelLeveraging,yu2026CharacterizingHigh}.

\subsubsection{Baselines}
\textbf{HBF-only} uses six 16-die all-Flash stacks per sparse-attention GPU, with FIFO reads and random per-plane writes. Our \textbf{HBFlex-style} baseline~\cite{zhong2026HBFlexFlexible} adapts hot-plane offload and placement-aware thread-block ordering to all-HBF stacks with 40~MiB of base-die SRAM per stack. We approximate its thread-block ordering with fixed-size chunks. Our \textbf{SPLASH-style} baseline~\cite{jonnalagadda2026SPLASHCoDesigning} adapts SPLASH's recent-token DRAM window and co-selection packing to HBFWeave's mixed stacks and DRAM budget. It keeps recent KV-cache pages in DRAM and older pages in HBF. SPLASH also caps query-scored page selections per plane to balance reads. \textbf{HBFWeave} uses mixed DRAM--HBF stacks with reuse-aware cache admission, ready-plane read scheduling, read-prioritized program waves, and request-grouped block allocation within each Flash plane, as described in Section~\ref{sec:design}.

\begin{table}[!t]
  \caption{HBFWeave system configuration.}
  \label{tab:system-configuration}
  \centering
  \footnotesize
  \begin{tabular}{@{}p{0.31\columnwidth}p{0.63\columnwidth}@{}}
    \toprule
    Parameter & Value \\
    \midrule
    Stack layout & 6 stacks/SA GPU; 16 memory dies/stack \\
    Flash organization & 32~GiB/die; 32 planes/die; 1,024 blocks/plane; 256 pages/block; 4~KiB/page \\
    Flash timing & Sense: 4~$\mu$s; program: 50~$\mu$s; erase: 2~ms \\
    DRAM organization & 2~GiB/die; 16 $\times$ 64-bit channels at 6~Gb/s/pin; 32 bank units/channel; 1~KiB rows \\
    DRAM timing (ns) & tCK=0.667, tCL=10, tRCDRD=20, tRP=16.7\newline tRAS=28.7, tRFC=400; tREFI=3.9~$\mu$s \\
    Base-die controller & 1,024 MSHRs; 256~KiB request queue; 8~MiB SRAM; 16 Flash channels/stack (0.125~$\mu$s command interval, 32 pending pages/channel) \\
    DRAM cache & 4~KiB cache pages; 16-way set associative \\
    SA GPU & FP8: 1.98~PFLOP/s; FP32: 66.9~TFLOP/s \\
    HC GPU & 288~GB HBM (22~TB/s); FP8: 17.5~PFLOP/s\\
    Interconnect & 900~GB/s SA--HC link/GPU \\
    \bottomrule
  \end{tabular}
\end{table}

\begin{table}[!t]
  \caption{GPU counts and DRAM capacity by case.}
  \label{tab:case-resources}
  \centering
  \scriptsize
  \setlength{\tabcolsep}{2pt}
  \begin{tabular*}{\columnwidth}{@{}l@{\extracolsep{\fill}}cl|lcl@{}}
    \toprule
    B/Seq & SA/HC & KV/index/reserve/WB & B/Seq & SA/HC & KV/index/reserve/WB \\
    \midrule
    Qwen & & & GLM & & \\
    64/128K  & 1/1 & 272/105/11/1,660 & 64/128K  & 1/2 & 208/202/7/1,631 \\
    64/1M    & 1/1 & 272/105/12/1,659 & 64/1M    & 1/2 & 208/203/9/1,628 \\
    256/128K & 1/1 & 1,040/105/40/863 & 256/128K & 1/2 & 784/202/26/1,036 \\
    256/1M   & 2/1 & 528/105/23/1,392 & 256/1M   & 1/2 & 816/203/32/997 \\
    \midrule
    DeepSeek & & & MiniMax & & \\
    64/128K  & 1/1 & 256/679/8/1,105 & 64/128K  & 1/2 & 1,184/384/38/442 \\
    64/1M    & 1/2 & 288/690/12/1,058 & 64/1M    & 2/3 & 608/380/27/1,033 \\
    256/128K & 1/1 & 1,008/679/31/330 & 256/128K & 2/2 & 2,352/357/77/1,310 \\
    256/1M   & 1/2 & 1,136/690/43/179 & 256/1M   & 8/8 & 608/380/27/1,033 \\
    \bottomrule
  \end{tabular*}
  \par\vspace{2pt}
  \noindent\parbox{\columnwidth}{\scriptsize MiB/stack. B/Seq: batch/context; SA/HC: GPUs; KV: cache target $C$; index: coarse summaries; reserve: mainly cache tags and history metadata; WB: write buffer. One DRAM die/stack except MiniMax 256/128K (two).}
\end{table}

\subsection{End-to-End Decode Performance}
Fig.~\ref{fig:e2e-comparison} compares HBFWeave with HBF-only, HBFlex-style, and SPLASH-style in 16 cases. Relative to HBF-only with random block placement, HBFWeave achieves 1.58--4.25$\times$ throughput (2.65$\times$ on average) and average P50/P90 TPOT speedups of 2.55$\times$/3.81$\times$. HBFlex-style and SPLASH-style average 1.38$\times$ and 1.47$\times$ throughput speedups, respectively. HBFWeave improves estimated energy efficiency by 2.12\(\times\) on average over HBF-only, compared with 1.51\(\times\) for HBFlex-style and 1.58\(\times\) for SPLASH-style. HBFWeave's read benefit comes from caching reused KV pages and issuing residual misses on ready planes, reducing both repeated Flash senses and waits behind busy planes. At $B=256$ and 128K, per-step sparse-read time falls from 11.9 to 2.4~ms for Qwen and from 22.7 to 8.4~ms for DeepSeek. HBFlex-style's 40~MiB SRAM per stack cannot retain all reused pages, so frequently reused pages still incur Flash reads; CTA reordering offers limited further benefit because HBF-only already stripes pages across planes. SPLASH-style's recent-token window can miss older reused pages. Co-selection packing reduces distinct Flash pages for DeepSeek and GLM, whose selection units can share a page, but not for Qwen and MiniMax, whose units occupy at least one full page. DRAM staging and read-prioritized program waves reduce write-induced read stalls and lower P90 TPOT. Request-grouped blocks reduce GC copying and Flash energy. For Qwen at $B=256$ and 128K, write amplification falls from 3.97$\times$ with HBF-only to 1.38$\times$ with HBFWeave.

\subsection{Mechanism Analysis}

\begin{figure}[!t]
  \centering
  \includegraphics[width=\columnwidth,trim=0 4pt 0 0,clip]{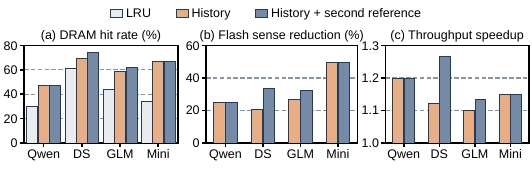}
  \caption{Cache admission at $B=256$, 128K: (a) hit rate; (b) Flash-sense reduction and (c) throughput speedup over LRU using 16-way DRAM caches.}
  \label{fig:cache-admission-ablation}
  \Description{Three panels compare LRU, historical-frequency admission, and its second-reference extension across four models in DRAM hit rate, percentage reduction in Flash senses relative to LRU, and serving throughput speedup over LRU.}
\end{figure}

\subsubsection{Cache Admission and Associativity}
Fig.~\ref{fig:cache-admission-ablation} compares LRU, historical-frequency admission, and its second-reference extension at $B=256$, 128K, using the cache budgets in Table~\ref{tab:case-resources}. All use 16-way caches and ready-plane scheduling. After 16 warm-up steps, we measure four consecutive decode steps across all sparse layers. The full policy raises DRAM hit rates from 30.1--61.4\% under LRU to 47.3--74.4\%, reducing Flash senses by 24.7--49.8\% and speeding up sparse reads by 1.41--1.52$\times$. Second-reference admission further helps DeepSeek and GLM, whose distinct KV units may share a Flash page, but has no effect on Qwen or MiniMax in these traces. The full policy improves throughput by 1.13--1.27\(\times\) over LRU in the serving replay. We use 16 ways throughout because hit rates level off beyond this point (Table~\ref{tab:cache-associativity}).

\begin{table}[!t]
  \caption{DRAM cache hit rate (\%) by associativity at $B=256$, 128K.}
  \label{tab:cache-associativity}
  \centering
  \scriptsize
  \setlength{\tabcolsep}{2pt}
  \begin{tabular*}{\columnwidth}{@{\extracolsep{\fill}}lrrrrrrr@{}}
    \toprule
    Model & 1 & 2 & 4 & 8 & \textbf{16} & 32 & 64 \\
    \midrule
    Qwen & 39.52 & 43.43 & 45.67 & 46.80 & \textbf{47.31} & 47.52 & 47.57 \\
    DeepSeek & 61.03 & 65.54 & 69.32 & 72.39 & \textbf{74.38} & 75.26 & 75.61 \\
    GLM & 47.11 & 53.17 & 57.57 & 60.53 & \textbf{62.08} & 62.66 & 62.89 \\
    MiniMax & 51.70 & 58.46 & 63.01 & 65.71 & \textbf{67.08} & 67.74 & 68.02 \\
    \bottomrule
  \end{tabular*}
\end{table}

\begin{figure}[!t]
  \centering
  \includegraphics[width=\columnwidth,trim=0 4pt 0 0,clip]{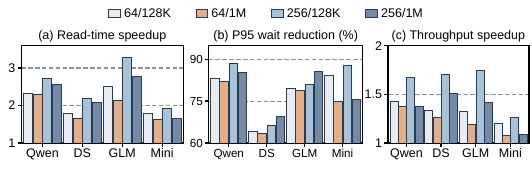}
  \caption{Ready-plane scheduling versus FIFO: (a) sparse-KV read speedup, (b) P95 miss-to-sense wait reduction, and (c) serving throughput speedup.}
  \label{fig:method2-read-scheduling}
  \Description{Three panels compare FIFO and ready-plane scheduling for four models at batch sizes 64 and 256 and contexts of 128K and 1M. They show sparse-KV read speedup, the reduction in P95 wait from tag completion to Flash sensing, and serving throughput speedup.}
\end{figure}

\subsubsection{Ready-Plane Scheduling}
Fig. 11 isolates ready-plane scheduling while holding all other settings fixed. It reduces P95 miss-to-sense wait by 63.5--88.5\%, speeding sparse-KV reads by 1.63--3.29$\times$ and serving throughput by 1.08--1.75$\times$. Larger batches supply more pending misses to ready planes, raising average read speedup from 2.02$\times$ at $B=64$ to 2.40$\times$ at $B=256$. Throughput speedup peaks at \(B=256\)/128K for all four models and is lower at 1M in every matched case. Larger batches increase the sparse-read share of TPOT and thus the throughput gain for DeepSeek and GLM.

\subsubsection{Write Interference and Flash Endurance}
\begin{figure}[!t]
  \centering
  \includegraphics[width=\columnwidth,trim=0 4pt 0 0,clip]{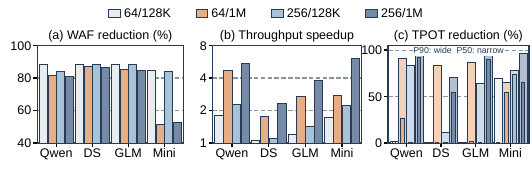}
  \caption{Write-path gains over direct random writes on matched mixed stacks: (a) WAF reduction, (b) throughput speedup, and (c) P50/P90 TPOT reduction.}
  \label{fig:method3-write-analysis}
  \Description{Three panels compare direct random writes with the complete HBFWeave write path for four models at batch sizes 64 and 256 and contexts 128K and 1M. They show percentage write-amplification reduction, throughput speedup, and percentage TPOT P50 and P90 reduction.}
\end{figure}

Fig.~\ref{fig:method3-write-analysis} compares HBFWeave's write policy with direct random writes on matched mixed stacks under the same read profile. HBFWeave lowers WAF by 51.2--88.6\% and raises throughput by 1.06--6.10\(\times\). Request grouping reduces GC copying while DRAM staging and program waves limit read stalls. Heavier prefill writes at 1M increase throughput gains for every model. HBFWeave limits intermittent program and GC stalls and thus often reduces P90 TPOT more than P50. It also reduces P50 when these stalls become frequent. P90 TPOT changes little for Qwen, DeepSeek and GLM but falls by 69.7\% for MiniMax at \(B=64\)/128K because its heavier KV-cache writes cause more interference under direct writes.

\subsection{Endurance and Sensitivity}\label{sec:endurance-sensitivity}
\subsubsection{Endurance} Across the 16 cases in Fig.~\ref{fig:e2e-comparison}, we compare HBFWeave with random-write HBF-only. Following~\cite{kim2026CapacityScalable}, we estimate Flash lifetime as \(T=C_{\mathrm{w}}E_{\mathrm{PE}}/(A_{\mathrm{w}}R_{\mathrm{w}}u)\). Each system uses its own writable capacity \(C_{\mathrm{w}}\) and replayed WAF \(A_{\mathrm{w}}\). Both assume \(E_{\mathrm{PE}}=100{,}000\) cycles, the same logical write rate \(R_{\mathrm{w}}\), continuous activity (\(u=1\)), and uniform wear. Fig.~\ref{fig:endurance-sensitivity}(a) shows an estimated 1.71--3.88\(\times\) Flash lifetime gain over HBF-only driven by reduced write amplification.

\subsubsection{Design Sensitivity} 
\begin{figure}[!t]
  \centering
  \includegraphics[width=\columnwidth,trim=0 4pt 0 0,clip]{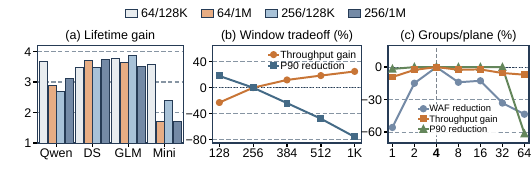}
  \caption{Endurance and design sensitivity: (a) lifetime gain; (b) Qwen window effects; (c) Qwen write-group effects on WAF, throughput, and TPOT.}
  \label{fig:endurance-sensitivity}
  \Description{Panel a shows projected HBFWeave lifetime relative to HBF-only for four models and four batch-context cases each. Panel b shows percentage throughput gain and P90 TPOT reduction for Qwen relative to a 256 MiB prefill program window as the window varies from 128 to 1024 MiB at batch size 256 and 1M context. Panel c shows Qwen WAF reduction, throughput gain, and P90 TPOT reduction relative to the four-group main setting as write groups per plane vary from 1 to 64 at batch size 256 and 128K context with a fixed 256 MiB window.}
\end{figure}

Fig.~\ref{fig:endurance-sensitivity}(b) shows that larger prefill windows raise throughput but increase P90 TPOT for Qwen at \(B=256\)/1M. Negative P90 reductions indicate TPOT above the 256~MiB baseline. Fig.~\ref{fig:endurance-sensitivity}(c) shows that four write groups per plane minimize WAF across 1--64 groups for Qwen at \(B=256\)/128K. Fewer groups mix more requests within blocks, whereas more groups can leave more blocks partially filled.

\subsubsection{Thermal Sensitivity} We derive power using replay and the energy parameters above. Following DeepStack~\cite{mo2026DeepStackFacilitating}, we add 70~W static power per GPU (10\% of H100 SXM's 700~W TDP~\cite{nvidiaH100SXM}). At 35~$^\circ$C ambient and $R_{\mathrm{sink}}=0.10$~K/W, HotSpot~\cite{huangHotSpotCompactThermal2006} estimates DRAM-die temperatures in Table~\ref{tab:thermal-sensitivity}, peaking at 68.7~$^\circ$C. No additional DRAM refresh overhead occurs because simulated die temperatures remain below the 85~\(^\circ\)C refresh threshold~\cite{ray2026ThermallyAwareSystemTechnology} in every case.

\begin{table}[!t]
  \caption{DRAM-die temperature ($^\circ$C) at 0.10~K/W.}
  \label{tab:thermal-sensitivity}
  \centering
  \footnotesize
  \renewcommand{\arraystretch}{0.82}
  \setlength{\aboverulesep}{1pt}
  \setlength{\belowrulesep}{1pt}
  \begin{tabular*}{\columnwidth}{@{}l@{\extracolsep{\fill}}rrrr@{}}
    \toprule
    $B$/Seq & Qwen & DeepSeek & GLM & MiniMax \\
    \midrule
    64/128K  & 49.9 & 49.8 & 48.7 & 49.9 \\
    64/1M    & 61.2 & 57.7 & 54.4 & 53.4 \\
    256/128K & 55.4 & 55.3 & 54.1 & 52.7 \\
    256/1M   & 62.1 & \textbf{68.7} & 67.9 & 53.4 \\
    \bottomrule
  \end{tabular*}
\end{table}

\section{Conclusion}\label{sec:discussion}

We present HBFWeave, a mixed DRAM--HBF system for long-context sparse-attention serving. It combines reuse-aware caching and ready-plane reads with read-prioritized program waves and request-grouped blocks. Across four LLMs and 16 cases, trace-driven evaluation shows average throughput and P50/P90 TPOT speedups of 2.65\(\times\) and 2.55\(\times\)/3.81\(\times\) over HBF-only. Estimated Flash lifetime gains of 1.71--3.88\(\times\) under uniform wear further support mixed DRAM--HBF stacks.

\bibliographystyle{IEEEtran}
\bibliography{refs}

\end{document}